\documentclass[pra,aps,amsmath,amssymb,eqsecnum,showkeys]{revtex4}
\newcommand{\pen}{\openone}
\newcommand{\iu}{{\mathtt{i}}}

\newcommand{\xdif}{{\mathrm{d}}}

\newcommand{\bsg}{{\boldsymbol{\sigma}}}
\newcommand{\hagam}{\hat{\Gamma}}
\newcommand{\am}{{\pmb{A}}}
\newcommand{\elm}{{\pmb{L}}}
\newcommand{\um}{{\pmb{U}}}
\newcommand{\rk}{{\mathrm{rank}}}
\newcommand{\av}{{\pmb{a}}}
\newcommand{\cbb}{{\mathbb{C}}}
\newcommand{\nbb}{{\mathbb{N}}}
\newcommand{\zbb}{{\mathbb{Z}}}
\newcommand{\wrh}{\widetilde{H}}
\newcommand{\wrn}{\widetilde{R}}
\newcommand{\zmx}{\mathbf{0}}
\newcommand{\clb}{{\mathcal{B}}}
\newcommand{\cle}{{\mathcal{E}}}
\newcommand{\clv}{{\mathcal{V}}}

\begin{document}
\clearpage
\preprint{}

\title{R\'{e}nyi and Tsallis entropies related to eigenfunctions of quantum graphs}

\author{Alexey E. Rastegin}
\affiliation{Department of Theoretical Physics, Irkutsk State University,
Gagarin Bv. 20, Irkutsk 664003, Russia}

\begin{abstract}
For certain families of finite quantum graphs, we study the
question of how eigenfunctions are distributed over the graph. To
characterize properties of the distribution, generalized entropies
of the R\'{e}nyi and Tsallis types are considered. The presented
approach is similar to entropic uncertainty relations of the
Maassen--Uffink type. Using the Riesz theorem, we derive lower
bounds on symmetrized generalized entropies of eigenfunctions. A
quality of such estimates will depend on boundary conditions used
at vertices of the given graph. R\'{e}nyi and Tsallis entropies of
eigenfunctions of star graphs are separately examined. Relations
between generalized entropies and variances of eigenfunctions are
considered as well. When such relations remain valid on average,
they may be used in studies of quantum ergodicity.
\end{abstract}


\maketitle

\pagenumbering{arabic}
\setcounter{page}{1}

\section{Introduction}\label{sec1}

Using quantum networks of one-dimensional wires to model physical
systems has a long history \cite{KS1999}. In the context of
quantum chaos on graphs, Kottos and Smilansky \cite{KS1997}
rediscovered the graph trace formula first discussed by Roth
\cite{roth83}. This result allows one to consider the connection
between random matrix theory and chaotic classical dynamics. In
such investigations, we try to understand the relationship between
quantum mechanics and classical chaos (see
\cite{kottos97,gnut06,smilansky07} for graph models and
\cite{stock99} for other topics). Quantum graphs are often used as
simplified models in mathematics, natural sciences, and
engineering \cite{berko12,berko16}. Such models are naturally
arisen in studies of nano- or meso-scale systems that are similar
to a neighborhood of a graph. Studies of differential operators on
metric graphs form an interesting branch of mathematical physics
\cite{exner97,kostr99,kuch04,kuch05,gnut05,kmn13}. Such operators
may be served as model systems arising in quantum chaos and
related questions of statistical physics. They are typically
treated as Hamiltonians including the negative Laplace
operator \cite{berko12}. Considering the Laplacian and other
operators on a metric graph, suitable boundary conditions at the
vertices should be assigned.

Entropies provide a powerful and flexible tool for investigation
of distribution properties. Such functions can often be used as
indicators of quantum chaos \cite{kz90}. The authors of
\cite{kameni14} used the standard Shannon entropy to characterize
eigenfunctions of quantum graphs. These results are inspired by
analogous studies of eigenfunctions on quantum maps
\cite{ananth07,gutkin10} and Riemannian manifolds
\cite{nonnen07,ananth08}. It is interesting that the notion of
quantum graphs {\it per se} can be approached through considering
one-dimensional piecewise linear maps \cite{kz01}. The approach of
\cite{kameni14} is to define entropies in terms of components of
the corresponding eigenvectors. For other purposes, this idea was
already realized in \cite{kz90}. The estimates of \cite{kameni14}
are essentially based on entropic uncertainty relations of the
Maassen--Uffink type \cite{maass}. Although the
Shannon entropy is fundamental, other entropic functions have
found use in various disciplines \cite{beck93,bengtsson}. The
R\'{e}nyi entropy \cite{renyi61} and the Tsallis entropy
\cite{tsallis} both give an important one-parameter extension of
the Shannon entropy. Generalized entropies provide an additional
tool in characterizing eigenfunctions on quantum graphs.

The aim of the present work is to characterize properties of
eigenfunctions on quantum graphs by means of the R\'{e}nyi and
Tsallis entropies. The paper is organized as follows. In Section
\ref{sec2}, we recall basic definitions concerning quantum graphs.
Section \ref{sec3} introduces R\'{e}nyi and Tsallis entropies
corresponding to an eigenfunction of a quantum graph. Further, we
obtain lower bounds on symmetrized generalized entropies of
eigenvectors associated with eigenfunctions on quantum graphs. We
also discuss circumstances under which the derived entropic bounds
may be useful. In Section \ref{sec4}, generalized entropies of
star graphs are considered. Due to a relative simplicity of such
graphs, their properties can usually be described with more
details. Section \ref{sec5} is devoted to relations between
entropies and the variance of a quantum graph. In particular, we
address an asymptotic behavior of averaged entropies, when the
number of bonds increases. In Section \ref{sec6}, we conclude the
paper with a summary of results obtained.

\section{Preliminaries}\label{sec2}

In this section, we recall definitions and introduce the notation.
The used notation closely follows chapter 1 of \cite{berko12}. Let
us consider a finite graph consisting of the set of vertices
$\clv=\{v\}$ and the set of undirected edges $\cle=\{e\}$. By
$V:=|\clv|$ and $E:=|\cle|$, we respectively mean the numbers of
vertices and edges. In the following, we will assume the absence
of loops and multiple edges. Two vertices $v$ and $v^{\prime}$ are
called adjacent, in symbols $v\sim{v}^{\prime}$, when there exists
an edge connecting them. For the given enumeration of vertices by
numbers $i\in\{1,2,\ldots,V\}$, each edge $e\in\cle$ is labeled by
pair $(ij)$ assumed to be symmetric. An undirected graph without
loops and multiple edges is fully specified by its $V\times{V}$
adjacency matrix with entries equal to $1$ for $(ij)\in\cle$ and
$0$ for $(ij)\not\in\cle$. The degree $d_{v}$ of a vertex $v$ is
the number of edges emanating from it. Graphs under consideration
are all assumed to be connected.

In the following, we will mainly deal with directed graphs. The
choice of orientation will be necessary for introducing
coordinates along edges. Now, each edge has one origin vertex and
one terminal vertex. Directed edges are referred to as
bonds and comprise the set of bonds $\clb=\{b\}$ of cardinality
$B=|\clb|$. Any undirected graph can be treated as a directed one
by assigning two bonds $b$ and $\bar{b}$ with opposite directions
to each edge $e$. In the following, we will deal only with such
digraphs. It will be convenient to use an enumeration of
vertices by numbers $i\in\{1,2,\ldots,V\}$. For the given
enumeration, each pair of adjacent vertices $i$ and $j$ is then
linked by the two bonds $b=[ij]$ and $\bar{b}=[ji]$. According to
\cite{smilansky07}, the notation $b=[ij]$ reads from the right to
the left so that $j$ is its origin and $i$ is its terminus. In
this case, we have $B=2E$ since for all $i\in\{1,2,\ldots,V\}$ the
number of incoming bonds equals the number of outgoing  ones. Up
to now, graphs were discussed from the combinatorial
perspective only as discrete structures. To approach quantum
graphs, edges should be considered as one-dimensional segments
sometimes called wires. Hence, digraphs will be equipped with
an additional structure that will make them metric graphs
\cite{berko12}.

A digraph becomes a metric graph, when each bond $b\in\clb$ is
assigned by a positive length $L_{b}\in(0,\infty)$ \cite{berko12}.
Thus, ordered points along $b$ are all identified with real
numbers between $0$ and $L_{b}$. On the bond $b=[ij]$, the
coordinate $0\leq{x}_{[ij]}\leq{L}_{b}$ is put by taking
$x_{[ij]}=0$ at the origin $j$ and $x_{[ij]}=L_{b}$ at the
terminus $i$. The lengths of the bonds that are reversed to each
other are treated as equal, namely $L_{b}=L_{\bar{b}}$. Hence, the
length $L_{e}$ of any edge $e\in\cle$ is also defined. Between the
coordinates along mutually reversed bonds, one has
\begin{equation}
x_{[ji]}=L_{e}-x_{[ij]}
\, , \label{xijx}
\end{equation}
where $e=(ij)$. The set of points of a metric graph include not
only its vertices, but all intermediate points on the edges
\cite{berko12}.

In order to consider quantum graphs, metric graphs should be
equipped with an additional operator called the Hamiltonian
\cite{berko12}. A function on the metric graph $\Gamma$ is
defined as a collection of $E$ scalar functions such that
$f_{e}:{\>}[0,L_{e}]\rightarrow\cbb$.
In the studies of quantum graphs, the most frequently used
operator is the negative second derivative acting on each edge.
Other physically important forms of the Hamiltonian are discussed
in \cite{berko12}. The definition of the quantum graph Hamiltonian
cannot be completed without adding smoothness conditions along the
edges and junction conditions at the vertices. Junction conditions
are similar to boundary ones used in the familiar case of
differential operators on a single interval.

In the case of the negative Laplace operator, the eigenvalue problem
is posed as \cite{berko12}
\begin{equation}
{}-\frac{\xdif^{2}f_{e}}{\xdif{x}_{e}^{2}}=\varkappa^{2}f_{e}(x_{e})
\, . \label{scprb}
\end{equation}
We usually look for real (positive) values $\varkappa\neq0$. With
the second derivative, we do not need to specify an orientation of
coordinates along the edges. This is required in other cases such
as the magnetic Schr\"{o}dinger operator \cite{berko12}. On each
edge $e=(ij)$, an eigenfunction with eigenvalue
$\varkappa^{2}\neq0$ is written in the form
\begin{equation}
f_{e}(x_{e})=a_{[ij]}\exp\bigl(\iu\varkappa\,x_{[ij]}\bigr)+a_{[ji]}\exp\bigl(\iu\varkappa\,x_{[ji]}\bigr)
\, . \label{eigfn}
\end{equation}
To complete the formulation, one imposes suitable boundary
conditions at the vertices where several edges meet. These
conditions should guarantee self-adjointness of the Hamiltonian
considered \cite{berko12,berko16}.

In general, boundary conditions can be described in two different
forms. In the first approach, certain pair of $d_{j}\times{d}_{j}$
matrices is assigned to $j$-th vertex, with
$j\in\{1,2,\ldots,V\}$. The second approach is posed by
prescribing how waves scatter at each vertex. This
approach is typical in studies of quantum chaos on graphs. It is
also more convenient for our purposes. The connections between the
two approaches are considered in section 2.1 of \cite{berko12}.
Boundary conditions can be specified in terms of unitary
scattering matrices assigned to graph vertices. For the
given vertex $j$ with degree $d_{j}$, the corresponding matrix
$\bsg^{(j)}$ has size $d_{j}\times{d}_{j}$ and entries
$\sigma^{(j)}_{[ij][jk]}$. At the vertex $j$, the boundary
conditions for eigenfunctions can be reformulated as
\begin{equation}
a_{[ij]}=
\sum_{k\sim{j}}
\sigma^{(j)}_{[ij][jk]}\exp\bigl(\iu\varkappa\,L_{[jk]}\bigr)\,a_{[jk]}
\, , \label{bocon}
\end{equation}
where the sum is taken over those vertices $k\in\{1,2,\ldots,V\}$
that are adjacent to $j$. The matrix $\bsg^{(j)}$ prescribes how
the vertex $j$ scatters waves incoming into it from adjacent
vertices. We will assume that vertex scattering matrices are
$\varkappa$-independent (for more details, see theorem 2.1.6 of
\cite{berko12}). The formula (\ref{bocon}) provides the
consistency requirement between the incoming and the outgoing
coefficients and must be true simultaneously at all the vertices
\cite{smilansky07}.

In general, the following two types of boundary conditions will
be used. According to the so-called Neumann conditions, the
function is continuous and the sum of its normal derivatives is
zero at each vertex \cite{kuch04,kuch05}. These conditions are
sometimes referred to as the Kirchhoff conditions
\cite{kostr99,kuch04} and the standard conditions \cite{berko12}.
Here, the scattering matrix at any vertex $j$ reads
\cite{tanner01}
\begin{equation}
\sigma^{(j)}_{[ij][jk]}=\frac{2}{d_{j}}-\delta_{ik}
\, , \label{neumn}
\end{equation}
where $\delta_{ik}$ is the Kronecker symbol. For graphs with large
degrees of vertices, the Neumann conditions imply a dominance of
back-scattering \cite{kameni14,tanner01}. For very large $d_{j}$,
the matrix $\!{}-\bsg^{(j)}$ will approach the identity matrix of the
corresponding size. The equi-transmitting boundary conditions were
introduced in \cite{harrison07}. The corresponding matrix elements
are characterized by the property \cite{harrison07}
\begin{equation}
\left|\sigma^{(j)}_{[ij][jk]}\right|^{2}=\frac{1-\delta_{ik}}{d_{j}-1}
\, . \label{eqtra}
\end{equation}
Thus, all the off-diagonal entries have equal amplitudes, and the
diagonal ones are zero. Hence, back-scattering is forbidden so
that an incoming wave is totally transmitted with equal weights to
outgoing bonds. These boundary conditions cannot be realized with
arbitrary $d_{j}$ \cite{harrison07}. The authors of
\cite{harrison07} gave examples of an explicit construction of
equi-transmitting scattering matrices. Their methods used
skew-Hadamard matrices \cite{white71,dokovic92,GKS202} and
properties of Dirichlet characters. In particular, the
orthogonality of Dirichlet characters is important here (see,
e.g., theorem 3.4 in chapter 5 of \cite{rose1994}). The second
construction provides an answer, in which $d_{j}-1$ is an odd
prime. Using the Legendre symbol as a Dirichlet character, one can
construct a symmetric equi-transmitting matrix with $d_{j}-1$
being a prime congruent to $1$ modulo $4$ \cite{harrison07}.

According to \cite{berko12}, quantum graphs are defined as metric
graphs equipped with a differential operator called the Hamiltonian and
accompanied by vertex conditions. That is, the
quantum graph $\hagam$ is a triple of metric graph $\Gamma$,
the Hamiltonian and boundary conditions in the form of matrices
$\bsg^{(j)}$ assigned to vertices $j\in\{1,2,\ldots,V\}$. In the
following, we restrict a consideration to the negative Laplace operator. For
each quantum graph $\hagam$, we write
the unitary evolution $B\times{B}$ matrix $\um_{\!\hagam}(\varkappa)$
with elements
\begin{equation}
u_{[ij][k\ell]}=\delta_{jk}\,\sigma^{(j)}_{[ij][j\ell]}\exp(\iu\varkappa\,L_{[j\ell]})
\, . \label{uijkel}
\end{equation}
Let $\av\in\cbb^{B}$ denote a column vector of coefficients
$a_{[ij]}$ that appear in (\ref{eigfn}). These coefficient
completely describe an eigenfunction of the problem (\ref{scprb}).
The consistency requirement (\ref{bocon}) then reduces to
\begin{equation}
\um_{\!\hagam}(\varkappa)\>\av=\av
\, . \label{umavv}
\end{equation}
To each eigenfunction, we can herewith assign an eigenvector of
$\um_{\!\hagam}(\varkappa)$ corresponding to eigenvalue $1$. This
vector specifies a distribution of the function (\ref{eigfn}) over
the graph. If $\varkappa^{2}$ is an eigenvalue of the problem
(\ref{scprb}), then $\varkappa$ obeys the secular equation
\begin{equation}
{\mathrm{det}}\bigl(\pen_{B}-\um_{\!\hagam}(\varkappa)\bigr)=0
\, , \label{umava}
\end{equation}
and {\it vice versa} \cite{berko12}. Many results on quantum
graphs hold under assumption that the eigenvalue is simple and the
eigenfunction is non-vanishing on vertices. As was shown in
\cite{fried2005,colin2015,berko2017}, these properties are generic
with respect to small perturbations of the edge lengths. Of
course, such perturbations have to break all symmetries of the
graph.

There are various ways to characterize eigenvectors of the unitary
evolution matrix. In the following, generalized entropies of the
R\'{e}nyi and Tsallis types will be utilized for such purposes.
When we associate entropic measures with finite structures, usual
vector norms in finite dimensions are convenient. For all
$p\geq1$, the usual vector $p$-norm of $B$-tuple is defined as
\begin{equation}
\|\av\|_{p}=\left(\sum\nolimits_{b=1}^{B}|a_{b}|^{p}\right)^{\!1/p}
 . \label{avpdf}
\end{equation}
The limiting value $p=\infty$ is allowed and leads to
$\max\{|a_{b}|:{\>}1\leq{b}\leq{B}\}$.

\section{Lower bounds on symmetrized entropies of eigenvectors}\label{sec3}

In this section, we derive lower bounds on symmetrized entropies
defined for an eigenfunction of some quantum graph. Let
$\zmx\neq\av\in\cbb^{B}$ and $0<\alpha\neq1$; then R\'{e}nyi's
$\alpha$-entropy of the column $\av$ with entries
$a_{1},a_{2},\ldots,a_{B}$ is defined as
\begin{equation}
R_{\alpha}(\av):=\frac{1}{1-\alpha}\>
\ln\!\left(\sum\nolimits_{b\in\clb} w_{b}^{\alpha}\right)
 ,  \label{renav}
\end{equation}
where the weights are put as
\begin{equation}
w_{b}=\|\av\|_{2}^{-2}\,|a_{b}|^{2}
\, . \label{wbdf}
\end{equation}
R\'{e}nyi considered this type of information measures in
connection with formal postulates characterizing entropic
functions \cite{renyi61}. The R\'{e}nyi $\alpha$-entropy cannot
increase with growth of $\alpha$ (see, e.g., section 5.3 of
\cite{beck93}). It has many interesting properties summarized in
section 2.7 of \cite{bengtsson}. The maximal value $\ln{B}$ of
(\ref{renav}) is reached when $w_{b}=1/B$ for all $b\in\clb$. The
following limiting cases should be mentioned separately. For
$\alpha\to\infty$, we have the min-entropy defined as
\begin{equation}
R_{\min}(\av):=-\ln(\max{w}_{b})
\, . \label{renmin}
\end{equation}
The limit $\alpha\to0$ gives the max-entropy. By $\rk(\av)$, we
denote the number of non-zero elements of $\av$. Then the
max-entropy is written as
\begin{equation}
R_{\max}(\av):=\ln\bigl\{\rk(\av)\bigr\}
\, . \label{renmax}
\end{equation}
As the R\'{e}nyi $\alpha$-entropy is a non-increasing function of
order $\alpha$, we have
\begin{equation}
R_{\min}(\av)\leq{R}_{\alpha}(\av)\leq{R}_{\max}(\av)
\, . \label{rmimx}
\end{equation}
For $\alpha=2$, the definition (\ref{renav}) gives the so-called
collision entropy.

Tsallis entropies form another especially important family of
generalized entropies. For $0<\alpha\neq1$, the Tsallis
$\alpha$-entropy of non-zero $\av\in\cbb^{B}$ is defined as
\begin{equation}
H_{\alpha}(\av):=\frac{1}{1-\alpha}\,
\left(\sum\nolimits_{b\in\clb} w_{b}^{\alpha}-1\right)
 .  \label{tsaav}
\end{equation}
With the factor $\left(2^{1-\alpha}-1\right)^{-1}$ instead of
$(1-\alpha)^{-1}$, this entropic form was considered by Havrda and
Charv\'{a}t \cite{havrda}. For $\xi>0$, we define the
$\alpha$-logarithm
\begin{equation}
\ln_{\alpha}(\xi):=
\begin{cases}
 \frac{\xi^{1-\alpha}-1}{1-\alpha}\>, & \text{ for } 0<\alpha\neq1\, , \\
 \ln\xi\, , & \text{ for } \alpha=1\,.
\end{cases}
\label{lnal}
\end{equation}
The maximal value of (\ref{tsaav}) is equal to $\ln_{\alpha}(B)$
and reached when $w_{b}=1/B$ for all $b\in\clb$.
The choice $\alpha=2$ gives the so-called linear entropy equal to
$1$ minus the sum of squared probabilities \cite{bengtsson}.
Conditional form of this entropy is directly related to the
minimal error probability on checking a finite or countable number
of hypotheses \cite{vajda68}. Due to non-additivity, the
Tsallis entropy is well known in non-extensive thermostatistics
\cite{tsallis}. Nevertheless, entropic functions of this type have
found use far beyond the context of thermostatistics. For
instance, such information measures were applied in formulation of
Bell inequalities \cite{rastann} and in studies of combinatorial
problems \cite{wei15,rastecount}. For Tsallis entropies, we do not
consider the limit $\alpha\to\infty$, as it leads to the same
zero value for all vectors. In the limit $\alpha\to1$, both the
above entropies reduce to the Shannon entropy
\begin{equation}
H_{1}(\av)=
-\sum\nolimits_{b\in\clb} w_{b}\,\ln{w}_{b}
\, .  \label{shaav}
\end{equation}
This entropy was used in studying properties of graph
eigenfunctions \cite{kameni14}.

In the following, entropic bounds will be expressed in terms of
the so-called symmetrized entropies. Such entropies were used in
formulating quantum-mechanical uncertainty relations
\cite{IBB06,rastmub}. Let positive orders $\alpha$ and $\beta$
satisfy $1/\alpha+1/\beta=2$. It is convenient to parametrize them
by means of $s\in[0;1)$,
\begin{equation}
\max\{\alpha,\beta\}=\frac{1}{1-s}
\ , \qquad
\min\{\alpha,\beta\}=\frac{1}{1+s}
\ . \label{paras}
\end{equation}
The symmetrized entropies R\'{e}nyi and Tsallis entropies are
respectively defined as
\begin{align}
\wrn_{s}(\av)&:=
\frac{1}{2}\,
\bigl( R_{\alpha}(\av)+R_{\beta}(\av)
\bigr)
\, , \label{syren}\\
\wrh_{s}(\av)&:=
\frac{1}{2}\,
\bigl( H_{\alpha}(\av)+H_{\beta}(\av)
\bigr)
\, . \label{sytsa}
\end{align}
The above condition on $\alpha$ and $\beta$ in entropic relations
will follow from the use of Riesz's theorem \cite{riesz27}. We
should also remember that the symmetrized entropies lead to the
standard Shannon entropy in the limit $s\to0$. For symmetrized
entropies of the R\'{e}nyi type, we will also use the limiting
value $s=1$, when
\begin{equation}
\wrn_{1}(\av)=
\frac{1}{2}\,
\bigl( R_{\min}(\av)+R_{\max}(\av)
\bigr)
\, . \label{syren1}
\end{equation}
For symmetrized entropies of the Tsallis type, the value $s=1$ is
not considered.

We shall now derive lower bounds on symmetrized R\'{e}nyi and
Tsallis entropies of eigenfunctions of quantum graphs. For
finite-dimensional systems, the entropic uncertainty principle is
most known in the formulation conjectured by Kraus \cite{kraus}
and later proved by Maassen and Uffink \cite{maass}. Their proof
is based upon a deep mathematical result known as Riesz's theorem
\cite{riesz27}. Using the Maassen--Uffink result, the authors of
\cite{kameni14} studied lower bounds on the Shannon entropy of
eigenfunctions on some quantum graphs. To examine the quantized
baker's map, the authors of \cite{ananth07} obtained lower bounds
on the entropies associated with semiclassical measures. In such
questions, the formulation of Maassen and Uffink is widely used.
So, we begin with recalling a simplified version of the Riesz
theorem.

It is sufficient to focus on unitary transformations
in finite dimensions. Due to the unitarity of $B\times{B}$ matrix
$\um=[[u_{bb^{\prime}}]]$, we have
\begin{equation}
\|\um\av\|_{2}=\|\av\|_{2}
\, . \label{ua2a}
\end{equation}
Using this fact, we can apply the Riesz theorem \cite{riesz27}
(see also theorem 297 of the book \cite{hardy}). Let positive
indices $p$ and $q$ be conjugated so that $1/p+1/q=1$, and let
$\eta$ be the maximal modulus of matrix entry, namely
\begin{equation}
\eta:=\max|u_{bb^{\prime}}|
\, . \label{cmudf}
\end{equation}
It then holds that, for $1\leq{q}\leq2$ and arbitrary
$\av\in\cbb^{B}$,
\begin{equation}
\|\um\av\|_{p}
\leq\eta^{(2-q)/q}\,\|\av\|_{q}
\, . \label{umpav}
\end{equation}
In general, inequalities of the form (\ref{umpav}) remain valid
for those transformations that do not increase the vector
$2$-norm. However, we deal with the unitary transformation, which
is invertible and the inversion is unitary as well. Under the same
conditions $1/p+1/q=1$ and $1\leq{q}\leq2$, we write a ``twin''
inequality
\begin{equation}
\|\av\|_{p}
\leq\eta^{(2-q)/q}\,\|\um\av\|_{q}
\, . \label{avpum}
\end{equation}
The latter is obtained by application Riesz's theorem to the
transformation $\um^{\dagger}$ acting on $\um\av$. The
above results hold irrespectively to the normalization of used
vectors.

Let positive indices $\alpha$ and $\beta$ obey
$1/\alpha+1/\beta=2$ and $\nu=\max\{\alpha,\beta\}$. For
$\av\in\cbb^{B}$ and unitary $B\times{B}$ matrix $\um$, we have
\begin{align}
R_{\alpha}(\av)+R_{\beta}(\um\av)
&\geq\!{}-2\ln\eta
\, , \label{abrmu}\\
H_{\alpha}(\av)+H_{\beta}(\um\av)
&\geq\ln_{\nu}\bigl(\eta^{-2}\bigr)
\, . \label{abhmu}
\end{align}
The relations (\ref{abrmu}) and (\ref{abhmu}) follow from
(\ref{umpav}) and (\ref{avpum}). The derivation uses a method
similar to that of \cite{rast104}. A reformulation for rank-one
resolutions of the identity in Hilbert space was later proposed in
\cite{rastmub}. The only distinction is that the papers
\cite{rastmub,rast104} deal with probabilistic vectors calculated
for a quantum state.

When $\av$ is an eigenvector of $\um$, the relations (\ref{abrmu})
and (\ref{abhmu}) involve two entropies of the same vector with
different entropic parameters. It will be convenient here to use
symmetrized entropies \cite{rastmub}. From (\ref{abrmu}) and
(\ref{abhmu}), we immediately obtain
\begin{align}
\wrn_{s}(\av)&\geq-\ln\eta
\, , \label{abrmu1}\\
\wrh_{s}(\av)
&\geq\frac{1}{2}\,\ln_{\nu}\bigl(\eta^{-2}\bigr)
\, , \label{abhmu1}
\end{align}
where $\nu=(1-s)^{-1}$. The formulas (\ref{abrmu1}) and
(\ref{abhmu1}) are one-parameter extensions of the inequality
\begin{equation}
H_{1}(\av)\geq-\ln\eta
\, . \label{abrmu0}
\end{equation}
The authors of \cite{kameni14} used (\ref{abrmu0}) for studying
the question of how eigenfunctions are distributed over a graph.
Some results were shown to be related to geometric properties of
the given graph. If $\av$ is an eigenvector of $\um$, it is also an
eigenvector of $\um^{t}$ for all $t\in\zbb$. In addition, the
power $\um^{t}$ is unitary as well. Hence, we have the following.

\newtheorem{prot1}{Proposition}
\begin{prot1}\label{pon1}
Let $\um$ be a unitary $B\times{B}$ matrix, and let $\av$ be an
eigenvector of $\,\um$. Denoting entries of $\,\um^{t}$ with natural
power $t$ by $u_{bb^{\prime}}^{(t)}$, we define
\begin{equation}
\eta^{(t)}=\max |u_{bb^{\prime}}^{(t)}|
\, . \label{cmudft}
\end{equation}
For all $t\in\nbb$ and $\nu=(1-s)^{-1}$, we then have
\begin{align}
 \wrn_{s}(\av)&\geq-\ln\eta^{(t)}
& (0\leq{s}\leq1)
\, , \label{abrmu11}\\
\wrh_{s}(\av)
&\geq\frac{1}{2}\,\ln_{\nu}\Bigl\{\bigl(\eta^{(t)}\bigr)^{-2}\Bigr\}
& (0\leq{s}<1)
\, . \label{abhmu12}
\end{align}
\end{prot1}

The maximal possible value of R\'{e}nyi entropies is $\ln{B}$.
Rescaling to the latter, we obtain lower bound on normalized
entropies,
\begin{equation}
\frac{\wrn_{s}(\av)}{\ln{B}}\geq\!{}-\frac{\ln\eta^{(t)}}{\ln{B}}
\ . \label{cmudft111}
\end{equation}
For $s=0$, this formula reduces to the lower bound on the
normalized Shannon entropy derived in \cite{kameni14}. The lower
bounds (\ref{abrmu11}) and (\ref{abhmu12}) will be useful, when
the quantity (\ref{cmudft}) is sufficiently far from $1$. Since
the squared absolute values of elements of the rows or columns of
a unitary matrix sum to one, they should not deviate essentially
from $1/B$. Thus, the above entropic relations may lead to a
good estimate, when some powers of the corresponding unitary
matrix are not too sparse. This condition can be treated in the context of stochastic
classical dynamics in a Markov chain \cite{kameni14}.

For quantum graphs with equi-transmitting boundary conditions, the
notion of graph girth is useful \cite{kameni14}. Let us take a
combinatorial graph without loops and multiple edges. Any sequence
$(v_{\tau},\ldots,v_{1},v_{0})$ of adjacent and distinct vertices
is referred to as the path of length $\tau$. If
$(v_{\tau-1},\ldots,v_{0})$ is a path and $\tau\geq3$, then the
sequence $(v_{0},v_{\tau-1},\ldots,v_{0})$ of adjacent vertices is
a $\tau$-cycle. The minimum length of a cycle contained in the
graph $\Gamma$ is its girth $g(\Gamma)$. For a quantum graph, we
always refer to the girth of underlying combinatorial structure.
Assigning two bonds to each edge, we will obtain two directed
paths from any undirected one. Recall also that vertices of a
regular graph are all of the same degree. Further, we consider
$(d+1)$-regular graphs with equi-transmitting boundary conditions.
The following statement holds.

\newtheorem{prot2}[prot1]{Proposition}
\begin{prot2}\label{pon2}
Let $\hagam$ be a $(d+1)$-regular quantum graph with
equi-transmitting boundary conditions and girth $g(\Gamma)$.
For each eigenvector $\av$ of $\,\um_{\!\hagam}(\varkappa)$ and
$\nu=(1-s)^{-1}$, we then have
\begin{align}
 \wrn_{s}(\av)&\geq\frac{g(\Gamma)}{4}\,\ln{d}
& (0\leq{s}\leq1)
\, , \label{gbrmu11}\\
\wrh_{s}(\av)
&\geq\frac{1}{2}\,\ln_{\nu}\bigl(d^{\>g(\Gamma)/2}\bigr)
& (0\leq{s}<1)
\, . \label{gbhmu12}
\end{align}
\end{prot2}

{\bf Proof.} Following \cite{kameni14}, we use a unitary matrix
$\um_{\!\hagam}(\varkappa)^{t}$, where the power $t$ obeys
$g(\Gamma)/2\leq{t}<g(\Gamma)/2+1$. By equi-transmitting
boundary conditions, back-scattering is forbidden. To a non-zero
entry $u_{bb^{\prime}}^{(t)}$, we can herewith assign a unique
path of length $\tau=t-1$ from the terminus of $b^{\prime}$ into
the origin of $b$. As our graph is $(d+1)$-regular, one finally
gets \cite{kameni14}
\begin{equation}
\bigl|u_{bb^{\prime}}^{(t)}\bigr|^{2}\leq{d}^{\,-g(\Gamma)/2}
\, . \label{ubbpg}
\end{equation}
Combining the latter with (\ref{abrmu11}) and (\ref{abhmu12})
completes the proof. $\square$

We will further consider a sequence of $(d+1)$-regular
combinatorial graphs $\Gamma_{n}$ with natural $n$ and suppose
that the number of vertices $V_{n}=|\clv_{n}|$ monotonically grows
with $n$. Such sequences of graphs are said to have large girth if
there exists a constant $C>0$ such that \cite{davidoff03}
\begin{equation}
g(\Gamma_{n})=\bigl(C+o(1)\bigr)\,\frac{\ln{V}_{n}}{\ln{d}}
 \label{ggnd}
\end{equation}
and $o(1)\to0$ for $n\to\infty$. In other words, the girth of a
sequence element increases proportionally to the logarithm of the
number of vertices. It can be shown that $C\leq2$ with necessity
\cite{davidoff03}. Erd\"{o}s and Sachs \cite{erdos63} gave a
non-constructive proof of the existence of large-girth families of
regular graphs with $C=1$. At the same time, explicit examples are
difficult to construct. The known explicit construction of such
families is due to Margulis \cite{margulis82}. In the papers
\cite{erdos63,margulis82}, undirected graphs are considered. They
can be transformed into digraphs as noticed above.

Let $\{\hagam_{n}\}$ be family of $(d+1)$-regular quantum graphs
with large girth and equi-transmitting boundary conditions, and
let $\av_{n}$ be an eigenfunction of
$\,\um_{\!\hagam_{n}\!}(\varkappa)$. Combining (\ref{gbrmu11}) with
(\ref{ggnd}) then gives
\begin{equation}
\frac{\wrn_{s}(\av_{n})}{\ln{V}_{n}}\geq\frac{C+o(1)}{4}
\ . \label{rubbpg}
\end{equation}
This formula extends one of the results of \cite{kameni14} to
symmetrized R\'{e}nyi entropies. As was mentioned in \cite{maass},
relations with a parametric dependence provide additional
possibilities to analyze distributions under consideration.

\section{Entropic characterization of eigenfunctions of star graphs}\label{sec4}

In this section, eigenfunctions of star graphs will be
characterized by means of entropies of associated eigenvectors.
Properties of eigenvalues and eigenfunctions of such graphs were
studied in detail
\cite{BK1999,BBK2001,exner01,winn03,berko04,gnut04}. A star
graph consists of a single central vertex together with outlying
vertices, each of which is connected only to the central one.
Here, the center has degree $E$ and all other vertices have degree
$1$, so that $V=E+1$. At the ends of the edges, Neumann boundary
conditions are used. In each end, an incoming wave is merely
reflected and will return into the center. Various boundary
conditions at the central vertex may be assumed. The spectrum and
eigenfunctions of a star graph do not behave typically for quantum
chaotic systems, when it has Neumann like conditions at the
central vertex \cite{BK1999,BBK2001}. For other scattering
matrices, e.g., equi-transmitting ones, the spectrum and
eigenfunctions do behave generically. We will denote vertices by
numbers $i\in\{0,1,\ldots,E\}$ with the central vertex $0$. In
this way, the edges are naturally labeled by elements of the set
$\{1,\ldots,E\}$. To $j$-th edge, with $j\in\{1,\ldots,E\}$, one
assigns the bonds $[j0]$ and $[0j]$. This notation mainly
coincides with \cite{winn03}.

Due to the structure of star graphs, their properties can often be
characterized in detail. Spectral determinants of such graphs were
considered in \cite{winn03}. An eigenfucntion on edge $e$ is
expressed as \cite{winn03}
\begin{equation}
\phi_{e}(x_{e})=A_{e}\cos\bigl(\varkappa\,(x_{e}-L_{e})\bigr)
\, . \label{phex}
\end{equation}
It is usually assumed that the coordinate $x_{e}$ is measured from
the central vertex, at which $x_{e}=0$. In general, the form
(\ref{phex}) holds for arbitrary boundary conditions at the
central vertex, using the Neumann conditions at the end of each
edge. To any eigenfunction, we can assign a column
$\am\in\cbb^{E}$ with entries $A_{1},A_{2},\ldots,A_{E}$. The
corresponding entropies are then calculated in line with the
definitions (\ref{renav}) and (\ref{tsaav}). The only point is
that we now deal with $E$ weights defined as
\begin{equation}
\varpi_{e}=\|\am\|_{2}^{-2}\,|A_{e}|^{2}
\, . \label{vpdf}
\end{equation}
Using (\ref{phex}), we can also construct an eigenfunction
corresponding to the form (\ref{eigfn}). Along $j$-th edge, where
$j\in\{1,\ldots,E\}$, we write
\begin{equation}
A_{j}\cos\bigl(\varkappa(x_{[j0]}-L_{j})\bigr)=
a_{[j0]}\exp\bigl(\iu\varkappa\,x_{[j0]}\bigr)+a_{[0j]}\exp\bigl(\iu\varkappa\,(L_{j}-x_{[j0]})\bigr)
\, , \label{eigfn0}
\end{equation}
where the coefficients of incoming and outgoing waves are,
respectively,
\begin{equation}
a_{[j0]}=\frac{1}{2}\,A_{j}\exp\bigl(-\iu\varkappa\,L_{j}\bigr)
\, , \qquad
a_{[0j]}=\frac{1}{2}\,A_{j}
\, . \label{coinot}
\end{equation}
For $j\in\{1,\ldots,E\}$, the matrix $\bsg^{(j)}$ has size
$1\times1$ and entry $1$. At the end of $j$-th edge, the condition
(\ref{bocon}) therefore reduces to multiplying $a_{[j0]}$ by the
factor $\exp\bigl(\iu\varkappa\,L_{j}\bigr)$. The latter coincides
with (\ref{coinot}).

Thus, we assign to a star graph the two columns $\av\in\cbb^{B}$
and $\am\in\cbb^{E}$ with entries related by (\ref{coinot}).
Hence, the following connection between weights takes place. If
the edge $e$ generates the bonds $b(e)$ and $\bar{b}(e)$, then
\begin{equation}
w_{b(e)}=w_{\bar{b}(e)}=\frac{\varpi_{e}}{2}
\, . \label{twbe}
\end{equation}
In the case of star graphs, we deal with two entropic functions
calculated on the base of $\{w_{b}\}$ and $\{\varpi_{e}\}$,
respectively. The link between theses functions is posed as
follows.

\newtheorem{prot3}[prot1]{Proposition}
\begin{prot3}\label{pon3}
Let columns $\av\in\cbb^{B}$ and $\am\in\cbb^{E}$ be assigned to
the given eigenfunction of a star graph. For
$\alpha\in[0,\infty]$, the two R\'{e}nyi $\alpha$-entropies are
related as
\begin{equation}
R_{\alpha}(\av)=R_{\alpha}(\am)+\ln2
\, . \label{twren}
\end{equation}
For $\alpha\in(0,\infty)$, the two Tsallis $\alpha$-entropies are
related as
\begin{equation}
H_{\alpha}(\av)=2^{1-\alpha}H_{\alpha}(\am)+\ln_{\alpha}(2)
\, . \label{twtsen}
\end{equation}
\end{prot3}

{\bf Proof.} Due to (\ref{twbe}), we first observe that
\begin{equation}
\sum\nolimits_{b\in\clb} w_{b}^{\alpha}=
\sum\nolimits_{e\in\cle} \left\{(w_{b(e)})^{\alpha}+(w_{\bar{b}(e)})^{\alpha}\right\}=
2^{1-\alpha}\sum\nolimits_{e\in\cle} \varpi_{e}^{\alpha}
\, . \label{twalp}
\end{equation}
Combining this with the definition (\ref{renav}) gives the claim
(\ref{twren}). Using (\ref{twalp}), the entropy $H_{\alpha}(\av)$
is represented as
\begin{equation}
\frac{1}{1-\alpha}\,
\left(2^{1-\alpha}\sum\nolimits_{e\in\cle} \varpi_{e}^{\alpha}-1\right)=
\frac{2^{1-\alpha}}{1-\alpha}\,\left(\sum\nolimits_{e\in\cle} \varpi_{e}^{\alpha}-1\right)
+\ln_{\alpha}(2)
\, . \label{alptw}
\end{equation}
The latter completes the proof of (\ref{twtsen}).
$\square$

In the case $\alpha=1$, both the formulas (\ref{twren}) and
(\ref{twtsen}) lead to the relation
\begin{equation}
H_{1}(\av)=H_{1}(\am)+\ln2
\, . \label{twshn}
\end{equation}
The latter was mentioned in \cite{kameni14}, but in the form with
normalized entropies. Rescaling by maximal entropic values is not
trivial, since the columns $\av$ and $\am$ have different numbers
of entries. According to (\ref{twren}), the R\'{e}nyi
$\alpha$-entropies of $\av$ and $\am$ differ by additive term
$\ln2$. The symmetrized entropies $\wrn_{s}(\av)$ and
$\wrn_{s}(\am)$ are linked by the same relation. It is not the
case for symmetrized Tsallis entropies due to specific factors of
the form $2^{1-\alpha}$. Rescaling by maximal entropic values, we
have
\begin{equation}
\frac{\wrn_{s}(\av)}{\ln{B}}=
\frac{\ln{E}}{\ln{E}+\ln2}\,\frac{\wrn_{s}(\am)}{\ln{E}}+\frac{\ln2}{\ln{E}+\ln2}
\ . \label{resren}
\end{equation}
To reach a stronger estimate, we should rather focus on normalized
R\'{e}nyi entropies of $\am$. In the case of star graphs, such
entropies are of primary interest. For normalized Shannon
entropies, this conclusion was formulated in \cite{kameni14}.

The Riesz theorem leads to lower bounds on various
entropies calculated with $\am$. The center is the only vertex
with complicated picture. The junction condition at the central
vertex is written in line with  (\ref{bocon}). Substituting
(\ref{coinot}) into this condition gives
\begin{equation}
A_{j}\exp\bigl(-\iu\varkappa\,L_{j}\bigr)
=\sum\nolimits_{k=1}^{E}\sigma_{jk}^{(0)}\exp\bigl(\iu\varkappa\,L_{k}\bigr)A_{k}
\, , \label{const0}
\end{equation}
where we accordingly shorten the notation of matrix entries,
$\sigma_{jk}^{(0)}\equiv\sigma_{[j0][0k]}^{(0)}$. In matrix form,
we write
\begin{equation}
\am=
\exp\bigl(\iu\varkappa\,\elm\bigr)\,\bsg^{(0)}\exp\bigl(\iu\varkappa\,\elm\bigr)\,\am
\, , \label{const1}
\end{equation}
with the diagonal $E\times{E}$ matrix
$\elm={\mathrm{diag}}\bigl(L_{1},L_{2},\ldots,L_{E}\bigr)$. In
other words, the column $\am$ is an eigenvector of certain unitary
matrix corresponding to eigenvalue $1$. Similarly to the formulas
(\ref{abrmu1}) and (\ref{abhmu1}), we have arrived at a
conclusion.

\newtheorem{prot4}[prot1]{Proposition}
\begin{prot4}\label{pon4}
Let column $\am\in\cbb^{E}$ be assigned to the given eigenfunction
of a star graph. Then we have
\begin{align}
\wrn_{s}(\am)&\geq-\ln\eta_{0}
& (0\leq{s}\leq1)
\, , \label{abrst}\\
\wrh_{s}(\am)&\geq\frac{1}{2}\,\ln_{\nu}\bigl(\eta_{0}^{-2}\bigr)
& (0\leq{s}<1)
\, , \label{abhst}
\end{align}
where $\nu=(1-s)^{-1}$ and $\eta_{0}:=\max|\sigma_{ee^{\prime}}^{(0)}|$.
\end{prot4}

To obtain further results, we should specify explicit conditions
at the central vertex. We first consider a star graph with the
Neumann boundary conditions at the central vertex. Suppose also
that $E\geq4$, whence $1-2/E\geq2/E$. In terms of symmetrized
entropies, one gives the relations
\begin{align}
\wrn_{s}(\am)&\geq-\ln\!\left(1-\frac{2}{E}\right)
& (0\leq{s}\leq1)
\, , \label{abrst4}\\
\wrh_{s}(\am)&\geq\frac{1}{2}\,\ln_{\nu}\!\left(\frac{E^{2}}{(E-2)^{2}}\right)
& (0\leq{s}<1)
\, , \label{abhst4}
\end{align}
where $\nu=(1-s)^{-1}$. For sufficiently large $E$, the
symmetrized R\'{e}nyi entropy obeys
\begin{equation}
\wrn_{s}(\am)\geq\frac{2}{E}+O\!\left(\frac{1}{E^{2}}\right)
 . \label{rasym4}
\end{equation}
Thus, we have extended one of the results of \cite{kameni14} to
two families of generalized entropies. With the Neumann boundary
conditions, an eigenfunction cannot be gathered on a single edge,
but may be concentrated on two edges only. The above lower bounds
coincide with the related discussion in subsection 4.1 of
\cite{kameni14}. For large $E$, these bounds become vanishing. As
was noted in \cite{kameni14}, the entropic uncertainty principle
cannot lead to a good bound for star graphs with Neumann like
conditions at the center.

Let us proceed to the equi-transmitting boundary conditions. For
equi-transmitting conditions at the central vertex, lower entropic
bounds turn out to be almost optimal. In this case, diagonal
entries of $\bsg^{(0)}$ are all zero. According to (\ref{eqtra}),
for $e\neq{e}^{\prime}$ we have
\begin{equation}
\bigl|\sigma_{ee^{\prime}}^{(0)}\bigr|^{2}=\frac{1}{E-1}
\ . \label{trast}
\end{equation}
Combining this with (\ref{abrst}) and (\ref{abhst}) gives the
lower bounds
\begin{align}
\wrn_{s}(\am)&\geq\frac{1}{2}\,\ln(E-1)
& (0\leq{s}\leq1)
\, , \label{trastre}\\
\wrh_{s}(\am)&\geq\frac{1}{2}\,\ln_{\nu}(E-1)
& (0\leq{s}<1)
\, . \label{trastts}
\end{align}
Of course, we assume here that $E>1$. For $s=0$, both the
relations reduce to the lower bound on the Shannon entropy. The
latter was proved in \cite{kameni14}, but in terms of normalized
entropies. Combining (\ref{twren}) with (\ref{trastre}), we also
obtain a good lower bound on R\'{e}nyi entropies of $\av$. For a
star graph with $E=B/2$ edges and equi-transmitting conditions at
the central vertex, it holds that
\begin{equation}
\frac{\wrn_{s}(\av)}{\ln{B}}\geq\frac{\ln(B-2)+\ln2}{2\ln{B}}
\ . \label{trastav}
\end{equation}
Asymptotically, this lower bound is close to $1/2$. At the same
time, the left-hand side of (\ref{trastav}) cannot exceed $1$. So,
the ratio of $\wrn_{s}(\av)$ to its maximal possible value may
range in sufficiently limited interval only. In the case of
equi-transmitting conditions at the central vertex, we have
obtained a good estimate of symmetrized entropies from below.

\section{Relations between generalized entropies and the variance}\label{sec5}

In this section, we will consider relations between entropies of
an eigenfunction and its variance. In this case, we do not need in
symmetrization of entropies with respect to the entropic
parameter. The variance is widely used to characterize how some
vector deviates from the equi-distributed one \cite{kameni14}. The
variance is defined as follows. Using the weights (\ref{wbdf}),
for the given $\av\in\cbb^{B}$ one has
\begin{equation}
D(\av):=\frac{1}{B}\,\sum\nolimits_{b\in\clb}(Bw_{b}-1)^{2}
\, . \label{dvard}
\end{equation}
When the given vector is equi-distributed, $w_{b}=1/B$ for all
$b\in\clb$ and $D(\av)=0$. When the variance is small, the vector
is close to equi-distribution. It will be more convenient to
represent the variance as
\begin{equation}
D(\av)=\!{}-1+B\sum\nolimits_{b\in\clb}w_{b}^{2}
\, . \label{rcold0}
\end{equation}
The maximal value of (\ref{dvard}) is therefore equal to $B-1$.
The authors of \cite{kameni14} showed how the variance is
connected with the concept of quantum ergodicity. In general,
quantum ergodicity is not realized on finite quantum graphs. It
should be replaced with weaker notion, which the authors of
\cite{gnut10} called the asymptotic quantum ergodicity. They
further developed a Gaussian random wave model on quantum graphs.
A version of this model was previously introduced in
\cite{gsw2004}. For ergodic systems the behavior of almost all
eigenfunctions in the semiclassical limit is described by the
quantum ergodicity theorem \cite{colin85,zelditch87}. One of
difficult problems here is to estimate the rate by which the
expectation values approach the classical mean
\cite{zelditch94,schub98,schub06,schub08}. Quantum graphs can be
used as a relatively simple model for studies such questions.
Various aspects of ergodicity on quantum graphs were addressed in
\cite{berko04,berko07,gnut08,masson15}.

First of all, the variance is directly connected with the collision
entropy. Combining (\ref{rcold0}) with (\ref{renav}) immediately
gives
\begin{equation}
R_{2}(\av)=\ln{B}-\ln\bigl(1+D(\av)\bigr)
\, . \label{rcold1}
\end{equation}
We also recall that the R\'{e}nyi $\alpha$-entropy cannot increase
with growth of $\alpha$. Normalizing R\'{e}nyi entropies by the
denominator equal to $\ln{B}$, for $\alpha\in[0;2]$ we obtain
\begin{equation}
\frac{R_{\alpha}(\av)}{\ln{B}}\geq1-\frac{\ln\bigl(1+D(\av)\bigr)}{\ln{B}}
\ . \label{rcold11}
\end{equation}
In particular, this lower bound is valid for the normalized
Shannon entropy. Since $\ln\bigl(1+D(\av)\bigr)\leq{D}(\av)$, the
result (\ref{rcold11}) has improved the statement of lemma 5 of
\cite{kameni14}.

A relation between the variance and R\'{e}nyi entropies of order
$\alpha>2$ is more complicated. Let us begin with the min-entropy
(\ref{rmimx}). For the probability distribution with $B$ weights
$w_{b}$, we have
\begin{equation}
\max\{w_{b}:{\>}b\in\clb\}\leq
\frac{1}{B}\left(1+\sqrt{(B-1)D(\av)}\,\right)
 . \label{mwbb}
\end{equation}
This result is obtained by combining (\ref{rcold0}) with lemma 3
of \cite{rastmub}. As the function $\xi\mapsto-\ln\xi$ decreases,
the min-entropy obeys
\begin{equation}
R_{\min}(\av)\geq\ln{B}-\ln\!\left(1+\sqrt{(B-1)D(\av)}\,\right)
 . \label{rcold2}
\end{equation}
Using the lower bounds (\ref{rcold1}) and (\ref{rcold2}), we
have arrived at a conclusion.

\newtheorem{prot5}[prot1]{Proposition}
\begin{prot5}\label{pon5}
For $\alpha\in[2,\infty]$, the R\'{e}nyi $\alpha$-entropy of an
eigenfunction is bounded from below as
\begin{equation}
R_{\alpha}(\av)\geq\ln{B}
-\frac{\ln\bigl(1+D(\av)\bigr)}{\alpha-1}
-\frac{\alpha-2}{\alpha-1}{\>}\ln\!\left(1+\sqrt{(B-1)D(\av)}\,\right)
 . \label{ag2lb}
\end{equation}
\end{prot5}

{\bf Proof.} It was proved in proposition 1 of \cite{rastosid}
that the R\'{e}nyi $\alpha$-entropy of order $\alpha\geq2$ obeys
\begin{equation}
R_{\alpha}(\av)\geq\frac{1}{\alpha-1}{\>}R_{2}(\av)+\frac{\alpha-2}{\alpha-1}{\>}R_{\min}(\av)
\, . \label{al2inf}
\end{equation}
Combining the latter with (\ref{rcold1}) and (\ref{rcold2})
completes the proof. $\square$

The relations (\ref{rcold11}) and (\ref{ag2lb}) may be used, when
we focus on mean values of some quantities. For a family of graphs
with finite spectral gap, a quantum ergodicity statement has been
formulated in \cite{gnut10}. Using this result, the authors of
\cite{kameni14} described a model, in which the variance of an
eigenfunction obeys on average $\langle{D}(\av)\rangle=O(1)$.
Let us consider an ensemble of eigenfunctions, over which
a mean value should be taken. Averaging (\ref{rcold11}), due to
convexity of the function $\xi\mapsto-\ln(1+\xi)$ we obtain
\begin{equation}
\frac{\langle{R}_{\alpha}(\av)\rangle}{\ln{B}}\geq1-\frac{\ln\bigl(1+\langle{D}(\av)\rangle\bigr)}{\ln{B}}
\ , \label{rcold11m}
\end{equation}
where $\alpha\in[0,2]$. Combining concavity of the square root
function with decreasing of $\xi\mapsto-\ln(1+\xi)$, for
$\alpha\geq2$ we write
\begin{equation}
\frac{\langle{R}_{\alpha}(\av)\rangle}{\ln{B}}\geq1
-\frac{\ln\bigl(1+\langle{D}(\av)\rangle\bigr)}{(\alpha-1)\ln{B}}
-\frac{\alpha-2}{(\alpha-1)\ln{B}}{\>}\ln\!\left(1+\sqrt{(B-1)\langle{D}(\av)\rangle}\,\right)
 . \label{ag2lbm}
\end{equation}
In both the formilas (\ref{rcold11m}) and (\ref{ag2lbm}), the
left-hand side cannot exceed $1$. We now suppose that
$\langle{D}(\av)\rangle=O(1)$ as $B\to\infty$. If the average of
variances goes to a constant, then the R\'{e}nyi entropy will tend
at a logarithmic rate to $1$. In \cite{kameni14}, this
claim was formulated and numerically supported for the Shannon
entropy. Our results show that similar reasons are applicable to
averaged R\'{e}nyi entropies.

Many findings of the above discussion can be reformulated with
entropies of the Tsallis type. Like (\ref{rcold1}), the
linear entropy is immediately connected with the variance.
Substituting $\alpha=2$ into (\ref{tsaav}) and using
(\ref{rcold0}), we get
\begin{equation}
H_{2}(\av)=1-\frac{1+D(\av)}{B}
\ . \label{tsaln}
\end{equation}
After averaging, we will still have the exact equality instead of
inequalities such as (\ref{rcold11m}) and (\ref{ag2lbm}). Assuming
$\langle{D}(\av)\rangle=O(1)$, the averaged linear entropy will
obviously tend to the limit $1$. For other entropic parameters, we
should separately consider the intervals $\alpha\in(0,2]$ and
$\alpha\in[2,\infty)$. The following statement holds.

\newtheorem{prot6}[prot1]{Proposition}
\begin{prot6}\label{pon6}
For $\alpha\in(0,2]$, the Tsallis $\alpha$-entropy of an
eigenfunction is bounded from below as
\begin{equation}
H_{\alpha}(\av)\geq\ln_{\alpha}\!\left(
\frac{B}{1+D(\av)}\right)
 . \label{a02ts}
\end{equation}
\end{prot6}

{\bf Proof.} According to (\ref{tsaav}) and (\ref{lnal}), the
Tsallis $\alpha$-entropy can be rewritten as
\begin{equation}
H_{\alpha}(\av)=\sum\nolimits_{b\in\clb}w_{b}\,\ln_{\alpha}\!\left(\frac{1}{w_{b}}\right)
 . \label{alnts0}
\end{equation}
By inspection of the second derivative, the function
$\xi\mapsto\ln_{\alpha}\bigl(1/\xi\bigr)$ is convex for
$\alpha\in(0,2]$. Due to Jensen's inequality, we then obtain
\begin{equation}
\sum\nolimits_{b\in\clb}w_{b}\,\ln_{\alpha}\!\left(\frac{1}{w_{b}}\right)\geq
\ln_{\alpha}\!\left\{\biggl(\sum\nolimits_{b\in\clb}w_{b}^{2}\biggr)^{\!-1}\right\}
 . \label{alnts1}
\end{equation}
Combinig the latter with (\ref{rcold0}) finally gives the claim
(\ref{a02ts}). $\square$

In a structure, the right-hand side of (\ref{a02ts}) is similar to
(\ref{rcold11}) multiplied by $\ln{B}$. Similarly to the R\'{e}nyi
case, we will consider averages over an ensemble of
eigenfunctions. Due to the mentioned concavity, averaging results
in
\begin{equation}
\langle{H}_{\alpha}(\av)\rangle\geq\ln_{\alpha}\!\left(
\frac{B}{1+\langle{D}(\av)\rangle}\right)
 , \label{a02ts2}
\end{equation}
where $0<\alpha\leq2$. Further, we recall the identity
\begin{equation}
\ln_{\alpha}(\xi{z})=\ln_{\alpha}(\xi)+\xi^{1-\alpha}\ln_{\alpha}(z)
\, , \label{ident}
\end{equation}
which follows from (\ref{lnal}) immediately. The maximal value of
Tsallis' $\alpha$-entropy with $B$ weights is equal to
$\ln_{\alpha}(B)$. Applying (\ref{ident}) to (\ref{a02ts2}) and
rescaling it by the denominator $\ln_{\alpha}(B)$,  for
$\alpha\in(0,2]$ we have
\begin{equation}
\frac{\langle{H}_{\alpha}(\av)\rangle}{\ln_{\alpha}(B)}\geq1
-\frac{1}{\ln_{\alpha}\bigl(1/B\bigr)}\,\ln_{\alpha}\!\left(\frac{1}{1+\langle{D}(\av)\rangle}\right)
 . \label{a02ts3}
\end{equation}
For $\alpha\geq1$, the modulus of $\ln_{\alpha}\bigl(1/B\bigr)$
tends to infinity as $B\to\infty$. When an ensemble is such that
$\langle{D}(\av)\rangle=O(1)$ in this limit, the ratio of averaged
Tsallis' $\alpha$-entropy to $\ln_{\alpha}(B)$ is bounded by $1$
from below. Also, this ratio cannot exceed $1$. We may conclude
that, on average, the Tsallis $\alpha$-entropies with
$\alpha\in(1,2)$ will reveal a behavior similar to both the
Shannon entropy and the linear entropy (\ref{tsaln}). These are
obtained from (\ref{tsaav}) for $\alpha=1$ and $\alpha=2$,
respectively.

For $\alpha\geq2$, lower bounds on the Tsallis $\alpha$-entropy
are more complicated in formulation. For such values of $\alpha$,
we first write the inequality
\begin{eqnarray}
\sum\nolimits_{b\in\clb} w_{b}^{\alpha}
&\leq
(\max{w}_{b})^{\alpha-2}\,\sum\nolimits_{b\in\clb} w_{b}^{2}
\nonumber\\
&\leq
B^{1-\alpha}\bigl(1+D(\av)\bigr)\left(1+\sqrt{(B-1)D(\av)}\,\right)^{\alpha-2}
\, , \label{anf1}
\end{eqnarray}
which is based on (\ref{mwbb}). For $\alpha\geq2$, the Tsallis
$\alpha$-entropy of an eigenfunction is bounded from below as
\begin{equation}
H_{\alpha}(\av)\geq\frac{1}{\alpha-1}
\left\{
1-B^{1-\alpha}\bigl(1+D(\av)\bigr)\left(1+\sqrt{(B-1)D(\av)}\,\right)^{\alpha-2}
\right\}
 . \label{anf2}
\end{equation}
However, converting the latter into a relation with averaged
values seems to be difficult. This question may deserve further
investigations.

To sum up, we see the following. Lower bounds on suitably rescaled
R\'{e}nyi and Tsallis entropies are expressed as $1$ minus a term
depending on the variance. Such normalized entropies are also
bounded from above by $1$. Many of the derived relations remain
valid after averaging over an ensemble of eigenfunctions. Suppose
that we deal with some eigenfunction ensemble and consider growing
number of bonds. When $\langle{D}(\av)\rangle=O(1)$ as
$B\to\infty$, averaged R\'{e}nyi's $\alpha$-entropy will tend at a
logarithmic rate to $1$ for all $\alpha\in[0,\infty]$. In the same
case, the ratio of averaged Tsallis $\alpha$-entropy to
$\ln_{\alpha}(B)$ will tend to $1$ for all $\alpha\in[1,2]$. These
findings suggested that, on average, an eigenfunction will be
distributed over a graph with equal weights.

\section{Conclusions}\label{sec6}

We have studied generalized entropies of eigenfunctions of finite
quantum graphs. The derivation is based on Riesz's theorem using
an approach similar to that of Maassen and Uffink \cite{maass}.
Lower bounds on the symmetrized R\'{e}nyi and Tsallis entropies
were given in terms of the maximal modulus among entries of the
unitary evolution matrix. Such estimates can sometimes be related
to certain geometric characteristics of a graph. Due to relative
simplicity of star graphs, they give a good example to test
derived entropic bounds. The central vertex of a star graph is the
only vertex, at which complicated redistribution of incoming waves
takes place. For star graphs with Neumann conditions at the
center, entropic lower bounds do not provide a good estimate. For
the case of equi-transmitting conditions at the center, our
approach leads to good lower bounds on symmetrized entropies.

Lower bounds for the R\'{e}nyi and Tsallis entropies without
symmetrization were expressed in terms of the variance of a graph
eigenfunction. The variance is a natural measure of the deviation
of corresponding eigenvector from the equi-distributed one. The
obtained bounds are shown to be useful, when we average quantities
over an ensemble of eigenfunctions and assume growing number of
bonds. If the considered ensemble is of small variation,
normalized entropies will tend to $1$. For the normalized Shannon
entropy, this fact was proved and illustrated in
\cite{kameni14}. We have extended the above conclusion to the
R\'{e}nyi $\alpha$-entropy for $\alpha\geq0$ and to the Tsallis
$\alpha$-entropy for $1\leq\alpha\leq2$. In models with growing
number of bonds and a limited average variance, we will see
the equi-distribution of an eigenfunction at the microscopic scale.

\end{document}